%% file: main.tex
\documentclass[10pt,conference]{IEEEtran}
\IEEEoverridecommandlockouts
\def\BibTeX{{\rm B\kern-.05em{\sc i\kern-.025em b}\kern-.08em T\kern-.1667em\lower.7ex\hbox{E}\kern-.125emX}}

\usepackage{settings}
\usepackage{listings}

\newboolean{9pager} 
\setboolean{9pager}{true}
\newif\ifshort
\ifthenelse{\boolean{9pager}}
{
\shorttrue 
}
{}

\begin{document}

\title{
Interference and Coverage Analysis of mmWave \\ Inter-Vehicle Broadcast with Directional Antennas 
\thanks{This work is supported in part by the NSF award 1827211.}
\ifthenelse{\boolean{9pager}}
{}
{ 
}
}

\author{
\IEEEauthorblockN{Tianyi Zhang, Hongwei Zhang, Zhibo Meng}
\IEEEauthorblockA{\textit{Department of Electrical and Computer Engineering, Iowa State University} \\
\{tianyiz,hongwei,zhibom\}@iastate.edu
}
}

\maketitle

\input{abstract.tex}
\input{introduction.tex}
\input{systemModel.tex}
\input{analysis.tex}
\input{result.tex}
\input{conclusion.tex}

\ifthenelse{\boolean{9pager}}
{}
{ \input{ack}  }

\bibliographystyle{IEEEtran}
\bibliography{references-DNC,references}

\end{document}

%% file: abstract.tex
\begin{abstract} \label{sec:abstract}
Thanks to the availability of large bandwidth and high-gain directional antennas at the millimeter-wave (mmWave) bands, mmWave communications have been considered as one of the primary solutions to meet the high data rates needs in vehicular networks. Unicast in mmWave vehicle-to-vehicle (V2V) communications has been well-studied, but much less attention has been paid to V2V broadcast which is required by many V2V applications such as active safety. 
    To fill the gap, this paper systematically investigates 
    mmWave V2V broadcast by considering the unique properties of mmWave signal propagation in V2V environments as well as the impacts of directional antennas and interference. Based on widely-accepted, high-fidelity system models, we mathematically analyze the receiver-side signal-to-interference-plus-noise-ratio (SINR) and broadcast coverage, and we study the impacts of blockage, inter-vehicle distance, vehicle density and beam pattern. 
Through comprehensive numerical analysis, we find out that, instead of a single unique optimal beamwidth, there exists an optimal range of beamwidth, in which the beamwidths have similar performance and can maximize the coverage. We also find out that the selection of carrier sensing range plays an important role as it highly influences the performance of the whole vehicular networks. 
    Our analysis provides unique insight into mmWave V2V broadcast, and it sheds light on designing effective V2V broadcast protocols. 
\end{abstract}


%% file: introduction.tex
\section{Introduction} \label{sec:Introduction}
\subHeading{Motivation \& Related work.}
Unicast in MmWave vehicular networks has been heavily investigated in the literature. In particular, Verdone \cite{verdone1997outage} studies the base station (BS) to vehicle communications at the 60GHz band and analyzes the outage probability. In \cite{wang2018mmwave}, the authors consider the interference from a nearby BS for vehicle-to-infrastructure communications and show that the interference mainly comes from the BSs on the same street. The article in \cite{petrov2018impact} characterizes the interference from the side lanes in both highway and urban scenarios. Based on measurements, the authors estimate the signal-to-noise ratio (SNR) and perform the capacity analysis. Giordani et al$.$ \cite{giordani2018coverage} present a stochastic model for characterizing the beam coverage and connectivity probability of a mmWave automotive network. The results show that stable connectivity requires both accurate beam alignment between vehicles and satisfactory signal quality.

Many investigations \cite{hisham2020adjacent, he2018enhanced, zhang2016research} 
have been carried out for vehicular networks in the sub-6GHz band, and studies such as  \cite{dai2006efficient,shen2006directional} have considered efficient broadcast using directional antennas. There have also been studies \cite{zhang2021adaptive,chen2019millimeter} that investigate mmWave broadcasting schemes for vehicle networks, but those studies do not consider the impact of interference. To the best of our knowledge, no study has focused on the performance of mmWave V2V broadcast with directional antennas and inter-vehicle interference. 
    Yet a basic requirement in vehicular networks is for a vehicle to share its status with close-by surrounding vehicles \cite{Zhang:V2X-survey}. To this end, V2V unicast with directional mmWave communications tends not to be the best solution when facing high mobility and the need of exchanging huge amount 
    of inertia, LiDAR, radar, video, and other sensing data, and broadcast can help vehicles exchange data more efficiently while lowering the chance of connection disruption due to beam misalignment.

\subHeading{Contributions.}
To gain insight into mmWave V2V broadcast, we perform the associated coverage and interference analysis 
and make the following contributions:
\begin{itemize}
    \item We analyze broadcast coverage and SINR in mmWave V2V networks with directional antennas using high-fidelity system models. We derive the equations to calculate receiver-side SINR and broadcast coverage, and we analyze the impacts of blockage, inter-vehicle distance, vehicle density, and beam pattern.
    
    \item We investigate the relation between the 
    mmWave beam pattern and broadcast coverage. Results show that there exists an optimal beamwidth range in which the beamwidths have similar performance and can maximize the coverage. The optimal beamwidth range varies with vehicle densities. The range is smaller when the vehicle network is sparser, so it's more critical to choose the right beamwidth in sparser networks. This result provides new insight into optimal beam pattern selection.
    
    \item We find that the broadcast coverage is subject to the selection of carrier sensing range. A too large or too small carrier sensing range lowers the network performance. Hence, there is an open problem of how to optimally select the carrier sensing range while taking into consideration the unique characteristics of mmWave, directionality of antennas, and high mobility of vehicles.
    
    \item We validate the analytical results through simulation studies.
\end{itemize}

\subHeading{Organization.}
The paper is organized as follows: Section~\ref{sec:SystemModel} introduces the system model. 
The analysis of blockage probability, SINR, and broadcast coverage 
is given in Section~\ref{sec:Analysis}, while the numerical and simulation results are discussed in Section~\ref{sec:Result}. We conclude with key findings in Section~\ref{sec:Conclusion}.


%% file: systemModel.tex
\section{System Model} \label{sec:SystemModel}

\subsection{V2V Broadcast Scenario}

\begin{figure}[!htb]
\vspace*{-0.1in}
\centering 
\includegraphics[width=0.4\textwidth]{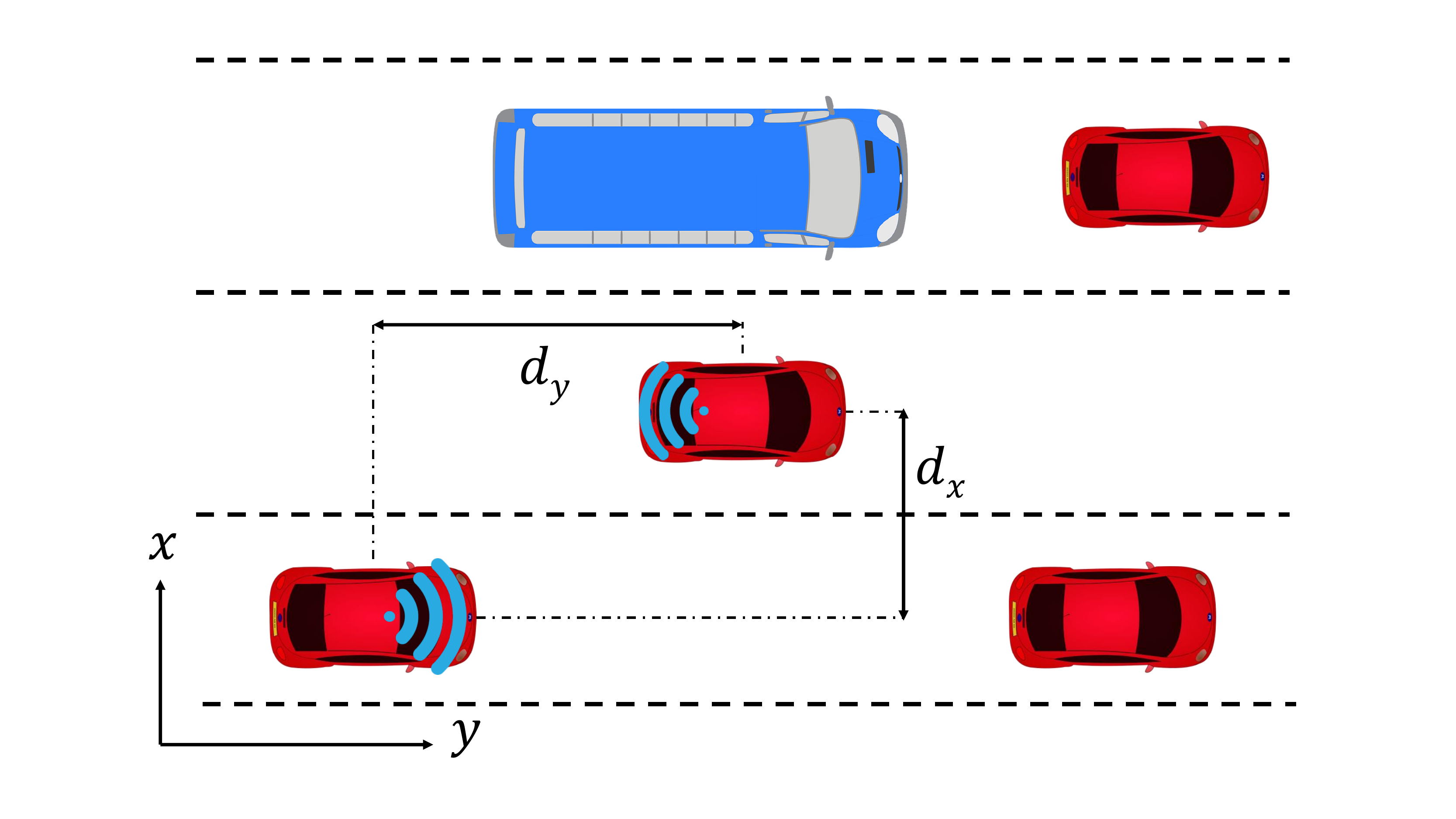}
\caption{Top view of V2V broadcast scenario}
     \label{scenario}
     \vspace*{-0.1in}
\end{figure}

We consider a 2D 3-lane road section with width $W$ as shown in Fig.~\ref{scenario}, 
where each vehicle periodically broadcasts its state (e.g., location and speed) to  neighboring vehicles. 
    This traffic model has been used by 3GPP to compare the performance of different technical options in the highway scenario \cite{3GPP},\footnote{Note that the analytical methodology and insights of this paper can be extended to other multi-lane scenarios.} 
and Guo et al$.$ \cite{guo2017spatial} have shown that, for free-flow vehicle traffic, the spatial distribution of vehicles along a highway lane is Poisson. Therefore, we assume that vehicles are  in the middle of their lanes and follow Poisson 
distributions, with an average spatial distribution density of $\lambda_i$ vehicles per meter along lane $i$. 
Like the setup in \cite{petrov2018impact,ozpolat2019connectivity}, each vehicle equips with the same antennas, including one directional antenna pointing to the dead front and another pointing to the dead astern. Both antennas are at the center of the rooftop. \footnote{Putting antennas at lower parts (e.g., bumpers) of vehicles does not work well for broadcast and hence is not considered in this study. 
} 
Similar to 3GPP Release 15 \cite{3GPP}, two vehicle types are defined as follows:
\begin{itemize}
    \item Passenger vehicle with high antenna position: length $l_1$, width $w_1$, height $h_1$, antenna height $h_{ant1}$.
    \item Truck/bus: length $l_2$, width $w_2$, height $h_2$, antenna height $h_{ant2}$.
\end{itemize}
The percentage of truck/bus among all the vehicles is $r_{tall}$.

\subsection{Path Loss Model}
Blockage effects play an important role in mmWave channels, and we adopt the following well-established 
log-distance path loss model \cite{yamamoto2008path,va2016millimeter}:
\vspace*{-0.05in}
\begin{small}
\begin{equation}
\begin{split}
PL_k=  10\alpha_{PL_k} \log_{10}{\sqrt{d_x^2+d_y^2}} & +\beta_{PL_k} \\
& +15\sqrt{d_x^2+d_y^2}/1000,
    \label{PL}
\end{split}
\vspace*{-0.07in}
\end{equation}\end{small}where $PL_k$ is the path loss between the transmitter and receiver in dB when there are $k$ blockers between them, 
$d_y$ is the y-axis displacement between the centers of the transmitter and receiver vehicles in the direction $y$ that is parallel to the vehicle moving direction, and $d_x$ is the x-axis displacement that is perpendicular to $y$ (see Fig. \ref{scenario}). $\alpha_{PL_k}$ and $\beta_{PL_k}$ are the model parameters with $\alpha_{PL_k}$ being the path loss exponent. Both $\alpha_{PL_k}$ and $\beta_{PL_k}$ vary with the number of blockers $k$ (e.g., see Table~\ref{tab2} in Section~\ref{sec:Result}). The last term in (\ref{PL}) is the atmospheric attenuation at 60 GHz.     One strength of the aforementioned model is that it is based on real-world measurement of mmWave vehicular communication channels, and all of its parameters can be instantiated based on the real-world measurement data from  \cite{yamamoto2008path,va2016millimeter}, thus ensuring the high-fidelity of our study. 

\subsection{Antenna Pattern}
\begin{figure}[!htb]
\vspace*{-0.1in}
\centering  
\label{Fig}
\includegraphics[width=0.4\textwidth]{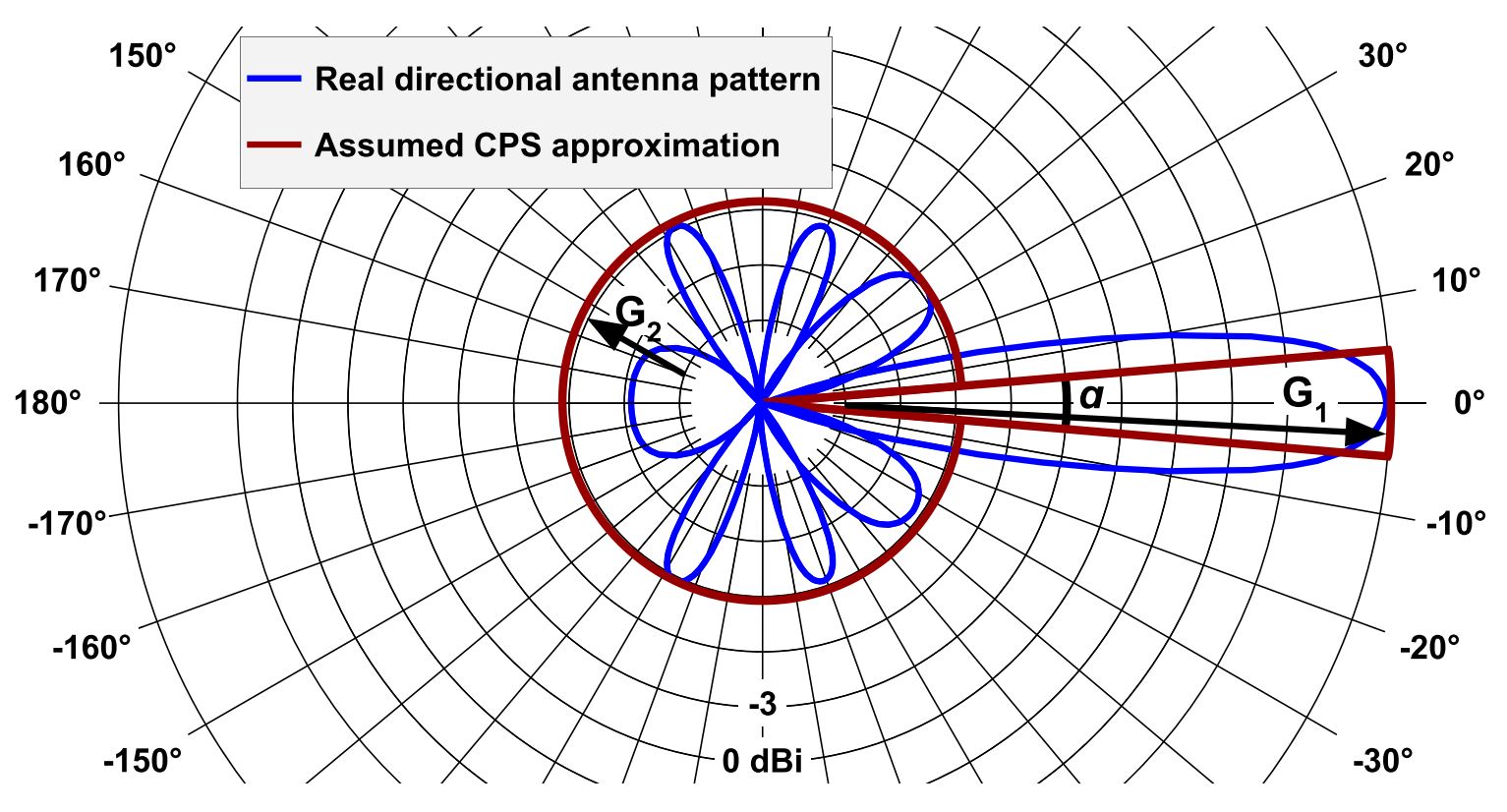}
\caption{{Cone-Plus-Sphere antenna pattern model}}
\label{figAntenna}
\vspace*{-0.1in}
\end{figure}

We adopt the well-established ``cone-plus-sphere" model \cite{petrov2017interference} to capture the antenna radiation pattern, as shown in Fig. \ref{figAntenna}. The antenna radiation pattern is modeled with 
a single cone-shaped beam as the main lobe, and a sphere around the antenna, which take into account the presence of side-lobes. Besides modeling directional communications, this model enables effective evaluation of the impact of side-lobes on interference, as shown in \cite{petrov2017interference}.  

$G_1$ and $G_2$ are the gains for the main and side lobes, given by
\vspace*{-0.05in}
\begin{equation}
    \begin{cases}
    G_1 = \frac{2}{1-\cos{(\alpha /2)}+k(1+\cos{(\alpha /2)})} \\
    G_2 = k G_1
    \end{cases},
    \vspace*{-0.03in}
\end{equation}
where $\alpha$ is the antenna directivity angle and $k \in (0,1)$ is the power ratio between the side lobe and the main lobe. In this model, the total transmit power doesn't change with the antenna beamwidth, so it allows us to compare the coverage and interference in different bandwidths in a fair condition. 

\subsection{Probability of transmission}

To reflect real-world situations, here we consider the MAC layer of IEEE 802.11bd, an enhanced V2X standard based on IEEE 802.11p. To characterize the channel access behavior, we adopt the model of semi-Markov process with absorbing state, 
which has been proposed in \cite{yin2014mac} to model V2V broadcast, 
instantiate the model with 802.11bd parameters as shown in Table~\ref{tab1} in Section~\ref{sec:Result}. 
In particular, given a node and all the other nodes within its carrier sensing range, we use the following two probabilities derived in \cite{yin2014mac}: 
\begin{itemize}
    \item probability $p_t$ that a node transmits at a time instant; 
    \item probability $p_c$ that more than one node 
    transmit concurrently.
\end{itemize} 
$p_t$ and $p_c$ capture the impacts of both carrier sensing and broadcast message intervals. Note that the impact of carrier sensing on the spatial distribution of concurrent transmitters and interference in 802.11bd is similar to that of interference-oriented scheduling in C-V2X \cite{CPS-TVT}. Thus the insight gained in this study also sheds light on C-V2X networks, even though the detailed study of C-V2X is beyond the scope of this paper. 

%% file: analysis.tex
\section{Interference and Coverage Analysis} \label{sec:Analysis}

\subsection{Blockage Probability} 

As vehicles are assumed to follow Poisson distributions along individual lanes, the distribution of trucks/buses also follows a Poisson distribution with an average spatial distribution density of $\lambda_i r_{tall}$ along lane $i$. \emph{When the transmitter and receiver are on different lanes}, blockers can be on the transmitter's and receiver's lane as well as any of the lanes in between.
For the lanes of the transmitter and receiver, the length of the area that the blocker can be placed on the y-axis is $d_{L1}=\frac{w_2}{2}\frac{d_y}{d_x}$. As the length on each lane is the same and considering the property of Poisson distributions, the total number of blockers in this case is the same as the case where all the blockers are on one lane and their spatial distribution density is the sum of the vehicle spatial distribution densities of the lanes of the transmitter and receiver. Thus the probability of having $k$ blockers on the same lane as the transmitter or receiver is
\vspace*{-0.05in}
\begin{equation} 
     P_{block,1}(k)=
             \frac{(r_{tall} d_{L1} (\lambda_{tx}+\lambda_{rx}))^k}{k!} e^{-r_{tall} d_{L1} (\lambda_{tx}+\lambda_{rx})}
    ,
    \vspace*{-0.05in}
\end{equation}
where $\lambda_{tx}$ and $\lambda_{rx}$ are the average spatial distribution densities of vehicles alone the lanes of the transmitter and receiver.

For the lanes between the transmitter and receiver, the length of the area that the blocker can be placed on the y-axis is $d_{L2}=w_2\frac{d_y}{d_x}+l_2$. The total number of blockers in this case is the same as the case where all the blockers are on one lane and their distribution density is $r_{tall} \sum_{i \in \mathcal{L}_2} \lambda_{i}$, with $\mathcal{L}_2$ denoting the lanes between that of the transmitter and receiver. The probability of having $k$ blockers on $\mathcal{L}_2$ is
\vspace*{-0.05in}
\begin{equation} 
     P_{block,2}(k)=
             \frac{(r_{tall} d_{L2} \sum_{i \in \mathcal{L}_2}  \lambda_i)^k}{k!} e^{-r_{tall} d_{L2} \sum_{i \in \mathcal{L}_2} \lambda_i}
    .
    \vspace*{-0.05in}
\end{equation}

\emph{When the transmitter and receiver are on the same lane}, blockers can only locate between the transmitter and the receiver. Hence the length of the area that the blocker can be placed on the y-axis is $d_y-l_2$. 

Thus the probability of having $k$ trucks/buses between a transmitter and receiver is
\vspace*{-0.04in}
\begin{small}
\begin{equation} 
     P_{tall,k}=
     \begin{cases}
             \sum_{i=0}^k P_{block,1}(i)P_{block,2}(k-i), & d_x \neq 0\\
            \frac{(r_{tall} (d_y-l_2)\lambda_{tx})^k}{k!} e^{-r_{tall} (d_y-l_2)
             \lambda_{tx}}, & d_x = 0
    \end{cases}
    .
     \label{blockages}
     \vspace*{-0.06in}
\end{equation}
\end{small}

Then, the probability that the signal from the transmitter to receiver is not blocked is
\vspace*{-0.05in}
\begin{equation}
    P_{b,0}=P_{tall,0}+(1-P_{tall,0})r_{tall}^2,
\vspace*{-0.05in}
\end{equation}
which includes the probabilities that 1) no truck/bus is between the transmitter and receiver,  2) there is at least one truck/bus in between but both the transmitter and receiver are trucks/buses so the signal isn't blocked. 
Similarly, the probability that the signal from the transmitter to receiver  is blocked by one blocker is
\vspace*{-0.05in}
\begin{equation}
    P_{b,1}=P_{tall,1}(1-r_{tall}^2).
    \vspace*{-0.05in}
\end{equation}
This happens when there is one truck/bus in between and the transmitter and receiver are not both trucks/buses.

\vspace*{-0.03in}
\subsection{Receiver-Side Interference}\label{secsinr}\
Given a transmitter and receiver with the x-axis and y-axis displacement of $d_x$ and $d_y$ respectively, when there are $k$ blockers, the received signal strength is as follows: 
\vspace*{-0.05in}
\begin{small}
\begin{equation}
     P_{r,k}(d_x,d_y)=\frac{P_t G_{t} G_{r}}{PL_k},
     \label{RxSignalStrengthk}
\vspace*{-0.05in}
\end{equation}
\end{small}where $P_t$ is the transmission power, $G_{t}$ and $G_{r}$ are the antenna gains at the transmitter and receiver side respectively, whose values are $G_1$ or $G_2$, depending on the relative positions $(d_x, d_y)$ of this pair of nodes. $PL_k$ is the path loss in magnitude. The received signal strength is 
\vspace*{-0.06in}
\begin{equation}
     P_{r}(d_x,d_y)=P_{b,0}P_{r,0}+P_{b,1}P_{r,1},
     \label{RxSignalStrength}
     \vspace*{-0.06in}
\end{equation}
which include the cases when there is no blocker or only 1 blocker. When there is more than one blocker, the received signal strength is too weak and can be ignored, which will be shown in section \ref{subImpactOfBlockage}.  
Here we assume that: 
\begin{itemize}
    \item All the transmitters have the same transmission powers. 
    \item All the transmitters and receivers have the same beam patterns, so the signals will be transmitted through either two main lobes or two side lobes.
    \item All the transmitters are using the front antennas while the receivers are using the rear antennas. This will reduce the interference and increase the probability that the transmitters and the receivers are communicating using the main lobes on both sides.
\end{itemize}

To analyze the impact of interference, we consider the MAC layer of IEEE 802.11bd. In 802.11bd, if carrier sensing is able to prevent concurrent transmissions by vehicles within the carrier sensing range $r_E$ from one another, then the minimum distance between concurrent transmitters would be the carrier sensing range. However, the nature of the CSMA/CA MAC of 802.11bd is such that, with probability $p_c$, vehicles within the carrier sensing range will transmit concurrently. Therefore, interference may be generated in two ways:
\begin{itemize}
    \item Primary interference $I_o$ that comes from concurrent transmitters $\mathcal{T}_o$ assuming that carrier sensing was able to prevent concurrent transmissions by vehicles within carrier sensing range from one another. 

    \item Secondary interference $I_c$ that comes from the concurrent transmitters $\mathcal{T}_c$ other than those causing $I_o$, that is, $\mathcal{T}_o \cap \mathcal{T}_c = \emptyset$. 
\end{itemize}

In what follows, we first analyze $I_o$. 
If without considering the carrier sensing mechanism, the interferers (i.e., concurrent transmitters $\mathcal{T}_o$) would follow a Poisson distribution with an average spatial distribution density of $\lambda_i p_t$ along each lane $i$. Thus the average distance between two neighbor interferers on the same lane is $\frac{1}{\lambda_i p_t}$. Considering the property of Poisson distribution, for one interferer, its nearest neighbor interferer can come from any one of the lanes, so the average distance between two neighbor interferers along the y-axis would be $\frac{1}{p_t \sum_{i \in \mathcal{L}_{All}} \lambda_i }$, where $\mathcal{L}_{All}$ denotes all the lanes and $\sum_{i \in \mathcal{L}_{All}}\lambda_i$ is the sum of the average spatial distribution densities on $\mathcal{L}_{All}$. 
With the carrier sensing mechanism working between interferers, however, other nodes are not allowed to transmit within the carrier sensing range $r_E$ of an interferer. Thus the carrier sensing range $r_E$ needs to be added to the distance between two neighbor interferers if without carrier sensing. The lane width $W$ is pretty small when compared with the distance between two neighbor interferers along the y-axis so we can ignore the effects caused by different lanes and use $r_E+\frac{1}{p_t \sum_{i \in \mathcal{L}_{All}}\lambda_i }$ as the average distance between two neighbor interferers. Then, the distance from the $k$th nearest interferer to the receiver can be approximated as 
\vspace*{-0.05in}
\begin{small}
\begin{equation}
     l_{I_k,R}=l_{S,R}+k(r_E+\frac{1}{p_t \sum_{i \in \mathcal{L}_{All}}\lambda_i }),
\vspace*{-0.05in}
\end{equation}
\end{small}where $l_{S,R}$ is the distance between the transmitter and the receiver.
    Accordingly, the total interference caused by the nodes outside the carrier sensing range is 
    \vspace*{-0.07in}
\begin{equation}
     I_o=\sum_{k=1}^\infty I_r(l_{I_k,R}),
     \label{IoSum}
     \vspace*{-0.07in}
\end{equation}
where $I_r(l_{I_k,R})$ is the received signal power from the $k$th nearest interferer to the receiver. $I_r(l_{I_k,R})$ is computed as
\begin{small}
\vspace*{-0.05in}
\begin{equation}
\label{eqn:I_r}
\begin{aligned}
     I_r(l_{I,R}) = & P_{0}P_r(0,l_{I,R}) + P_{W}P_r(W,\sqrt{l_{I,R}^2-W^2}) +   \\
     & P_{2W} P_r(2W,\sqrt{l_{I,R}^2-(2W)^2}), 
\end{aligned}
\vspace*{-0.05in}
\end{equation}\end{small}\noindent and it takes into account the fact that both the transmitter and receiver have equal probabilities of being on different lanes. $P_0=\frac{1}{3}$ is the probability that the x-axis distance between the transmitter and the receiver is 0, which means they are on the same lane; $P_W=\frac{4}{9}$ is the probability that the x-axis distance between the transmitter and the receiver is the lane width $W$, which means they are on the neighbor lanes; $P_{2W}=\frac{2}{9}$ is the probability that the x-axis distance between the transmitter and the receiver is $2W$, which means there is one lane between the lanes of the transmitter and receiver.

Now, we analyze the secondary interference $I_c$. The interferers $\mathcal{T}_c$ that generate the secondary interference are not influenced by carrier sensing and follow a Poisson distribution with an average spatial distribution density of $\lambda_i p_c$ along lane $i$. Accordingly, similar to the analysis of $I_o$, the total secondary interference is \vspace*{-0.07in}
\begin{small}
\begin{equation}
     I_c=\sum_{k=1}^\infty I_r(\frac{k}{p_c \sum_{i \in \mathcal{L}_{All}}\lambda_i }), 
     \label{IcSum}
     \vspace*{-0.05in}
\end{equation}
\end{small}
where the function $I_r(.)$ is as defined in (\ref{eqn:I_r}). 

\vspace*{-0.03in}
\subsection{Broadcast Coverage} 
\label{pes}

Inter-vehicle broadcast needs to be reliable for mission-critical applications such as active safety, thus the receiver-side SINR needs to be above a certain threshold $\gamma_{th}$ to ensure a reliable broadcast. To capture the performance of a V2V broadcast, we define the \emph{broadcast coverage} as the number of receivers whose SINRs are no less than a threshold $\gamma_{th}$. Besides capturing the number of receivers that can reliably receive a broadcast message, broadcast coverage reflects the size of the area in which a broadcast is reliable, which can be computed as the number of receivers divided by the vehicle distribution densities along traffic lanes. 
To decide the SINR threshold $\gamma_{th}$ to be used in this study, we consider a target packet error rate (PER) of $1 \times 10^{-5}$ as suggested by the ITU \cite{series2017minimum}. Then, from \cite{ma2021sinr}, we derive the required SINR as 23dB for 802.11bd with 256QAM and 5/6 coding rate.

With the interference formulas (\ref{IoSum}) and (\ref{IcSum}), we can derive as follows the minimum required receiver-side signal power, denoted by $P_{r,th}$, in order for a receiver to be considered ``within a broadcast coverage" (i.e., having an SINR no less than $\gamma_{th}$):
\vspace*{-0.05in}
\begin{equation}
     P_{r,th}=(N+I_o+I_c)\gamma_{th},
\vspace*{-0.05in}
\end{equation}
where $N=-174+10log_{10} B +NF$ is the noise, $B$ is the bandwidth, $NF$ is the noise figure.

Then, we can calculate the farthest distance $D_{r,th}$ from a transmitter to its receiver such that the receiver-side signal power is no less than the threshold $P_{r,th}$ using the inverse function of Equation~\ref{RxSignalStrength}. In particular, assuming $d_x = k \times W$, $k = 0, 1, 2$, (i.e., with $k$ lanes separation on road), $d_y$ can be derived and we denote it as $P_{r,k}^{-1}(P_{r,th})$. 

Then, counting all the receivers within $D_{r,th}$ distance from the transmitter on all the lanes, the expectation of the number of receivers whose SINRs are no less  than a threshold $\gamma_{th}$ can be expressed as follows:
\vspace*{-0.08in}
\begin{small}
\begin{equation}
\begin{aligned}
      E[N]= & \frac{1}{3} \lambda_1 (P_{r,0}^{-1}(P_{r,th})+P_{r,1}^{-1}(P_{r,th})+P_{r,2}^{-1}(P_{r,th})) + \\
     &  \frac{1}{3} \lambda_2(P_{r,0}^{-1}(P_{r,th})+2P_{r,1}^{-1}(P_{r,th})) 
     + \\
     &  \frac{1}{3} \lambda_3 (P_{r,0}^{-1}(P_{r,th})+P_{r,1}^{-1}(P_{r,th})+P_{r,2}^{-1}(P_{r,th})). 
\end{aligned}
\vspace*{-0.14in}
\end{equation}
\end{small}

%% file: result.tex
\section{Numerical and Simulation Results} \label{sec:Result}

Here we first perform numerical analysis in specific real-world settings, and then validate the analytical insight via simulation studies. 
    In particular, we consider a typical scenario suggested by the 3GPP release 15 \cite{3GPP} and papers \cite{va2016millimeter, perfecto2017millimeter}. Unless mentioned otherwise, the values of parameters used in our analysis are as shown in Tables~\ref{tab1}, \ref{tab2} and \ref{tab3}. 
Note that the key objective of this study is to distill insight into the behavior of mmWave inter-vehicle broadcast with directional antennas; the insight will help develop effective broadcast protocols, but the detailed study of specific protocols is beyond the scope of this work. 

\begin{table}[!htbp]
\vspace*{-0.05in}
\centering
\begin{tabular}{ |p{4.9cm}|p{2.7cm}| }
 \hline
 Frequency $f$ & 60GHz  \\
 \hline
 Transmission power $P_t$ & 23dBm \\
 \hline
 Beamwidth $\alpha$ & 30 $^{\circ }$ \\
 \hline
 Side-main lobe ratio $k$ & 0.1 \\
 \hline
 Carrier sensing range $r_E$ & 50m \\ 
 \hline
 Bandwidth $B$ &  20MHz \\
 \hline
 Noise figure $NF$ &  6dB \\
 \hline
 Passenger vehicle size $l_1 \times w_1 \times h_1$ & 5m$\times$2m$\times$1.6m \\
 \hline
 Passenger vehicle antenna height $h_{ant1}$ & 1.6m \\
 \hline
 Truck/bus size $l_2 \times w_2 \times h_2$ & 13m$\times$2.6m$\times$3m \\
 \hline
 Truck/bus antenna height $h_{ant1}$ & 3m \\
 \hline
 Lane width $W$ & 3.2m \\
 \hline
 Number of lanes $n$ & 3 \\
 \hline
 Percentage of bus/truck $r_{tall}$ &  0.1 \\
 \hline
 Packet generation interval  &  0.1s \\
 \hline
 Packet length &  200bytes \\
 \hline
 Symbol duration &  6.4$\mu$s \\
 \hline
 Transmission latency &  100$\mu$s \\
 \hline
\end{tabular}
\caption{System parameters}
\label{tab1}
\vspace*{-0.25in}
\end{table}

\begin{table}[!htbp]
\centering
\begin{tabular}{ |p{2cm}||p{1cm} p{1cm} p{1cm}| }
 \hline
   & Lane 1 & Lane 2 & Lane 3\\
 \hline
 Low & 0.05 & 0.07 & 0.10 \\
 Intermediate & 0.07 & 0.10  & 0.13 \\
 High & 0.09 & 0.13  & 0.17 \\
 \hline
\end{tabular}
\caption{Vehicle density (vehicles/m) in different traffic conditions}
\label{tab2}
\vspace*{-0.25in}
\end{table}


\begin{table}[!htbp]
\centering
\begin{tabular}{ |p{2cm}||p{1cm} p{1cm} p{1cm}| }
 \hline
 \# blockers  & 0 & 1 & 2\\
 \hline
 $\alpha_{PL}$ & 1.77 & 1.71 & 0.635 \\
 $\beta_{PL}$ & 70 & 78.6  & 115 \\
 \hline
\end{tabular}
\caption{Path loss model parameters}
\label{tab3}
\vspace*{-0.3in}
\end{table}

\begin{figure*}[!tb] 
\vspace*{-0.1in}
\figSpacing
\begin{minipage}[t]{0.321\linewidth}
\centering
\setlength{\abovecaptionskip}{0.cm}
\includegraphics[width=\textwidth]{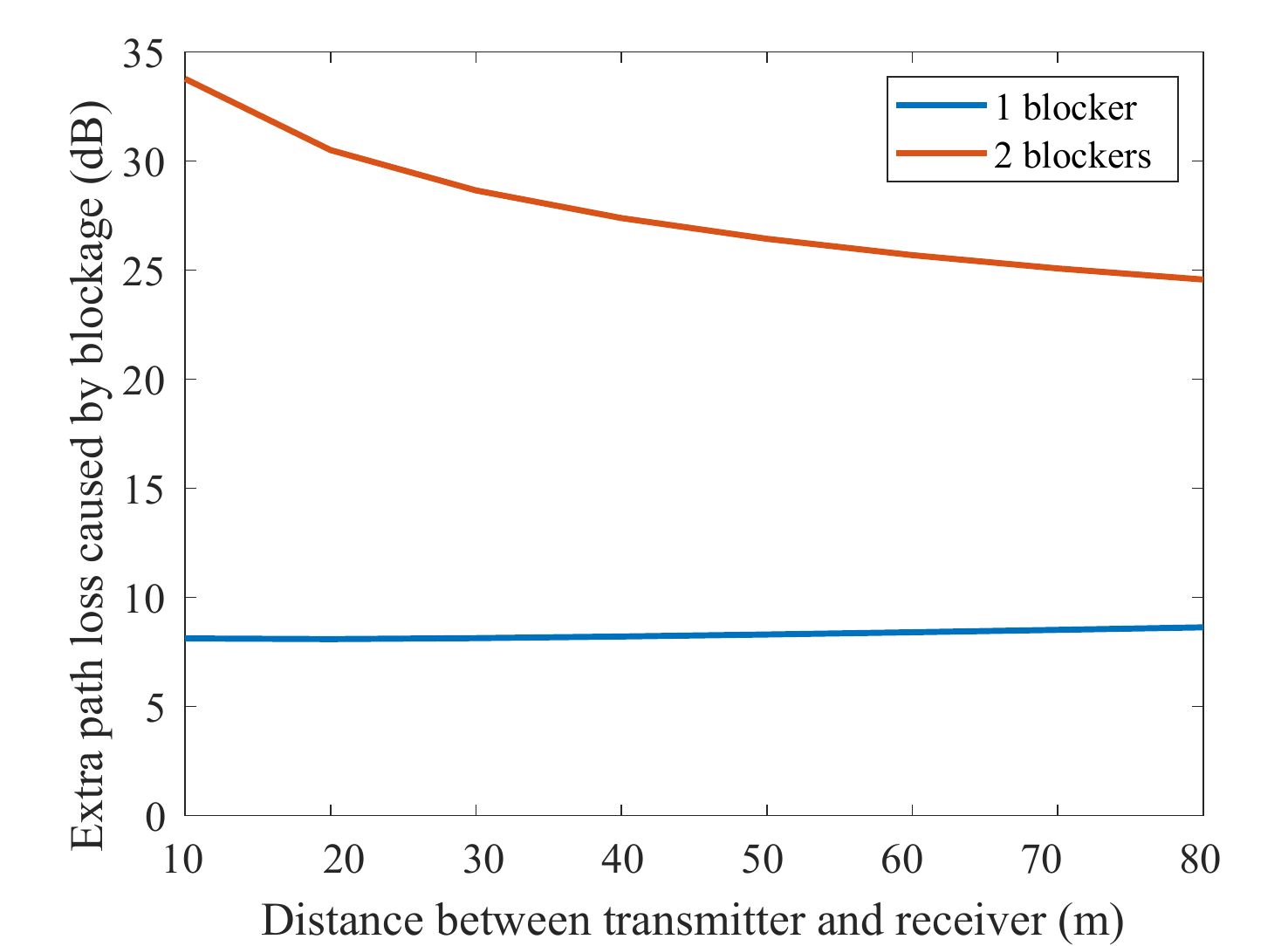}
\caption{The impact of different numbers of blocker} 
\label{blockage}
\end{minipage}	
\figSpacing
\begin{minipage}[t]{0.321\linewidth}
\centering
\setlength{\abovecaptionskip}{0.cm}
\includegraphics[width=\textwidth]{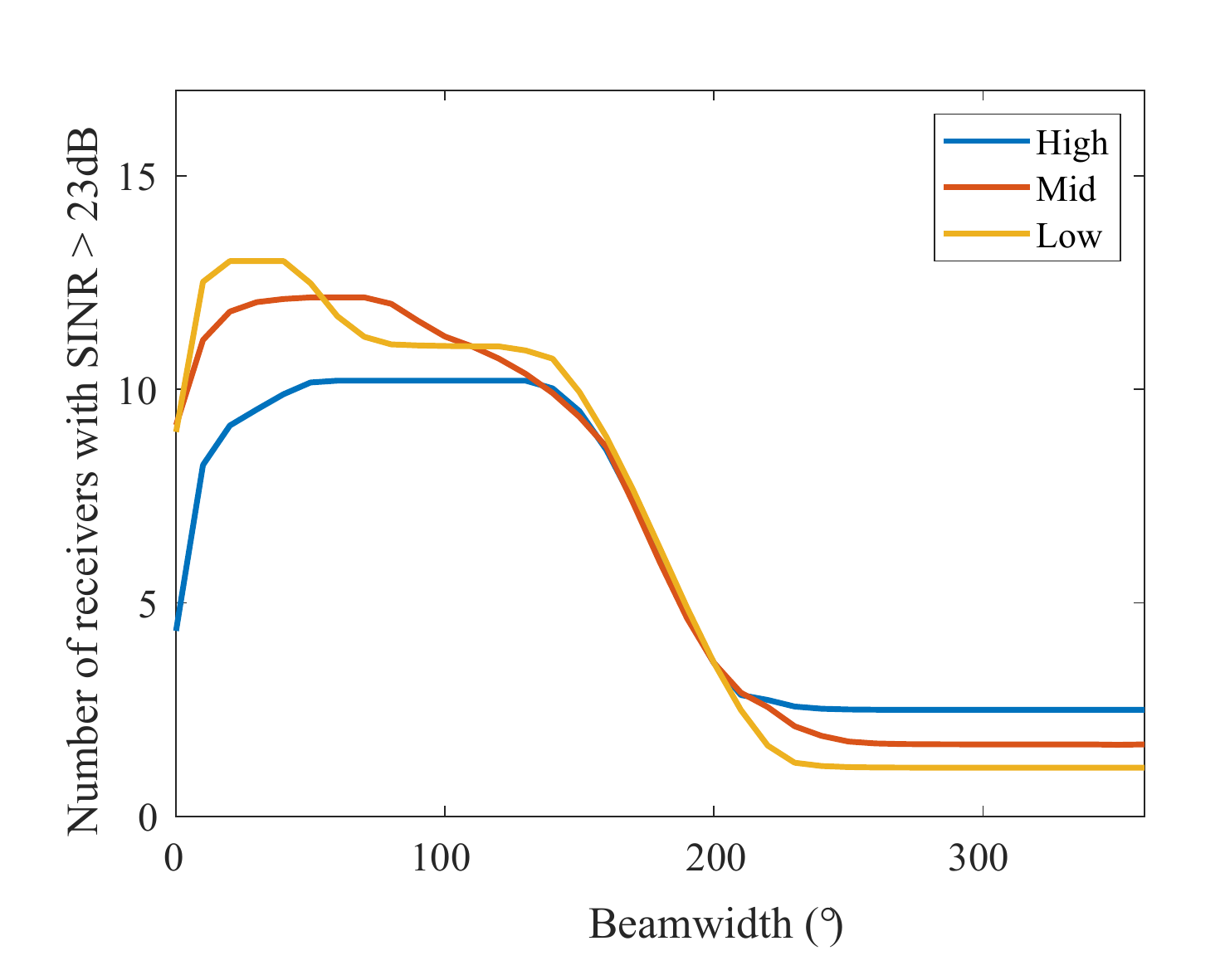}
\caption{The impacts of beamwidth and vehicle density} \label{density}
\end{minipage}	
\figSpacing
\begin{minipage}[t]{0.321\linewidth}
\centering
\setlength{\abovecaptionskip}{0.cm}
\includegraphics[width=\textwidth]{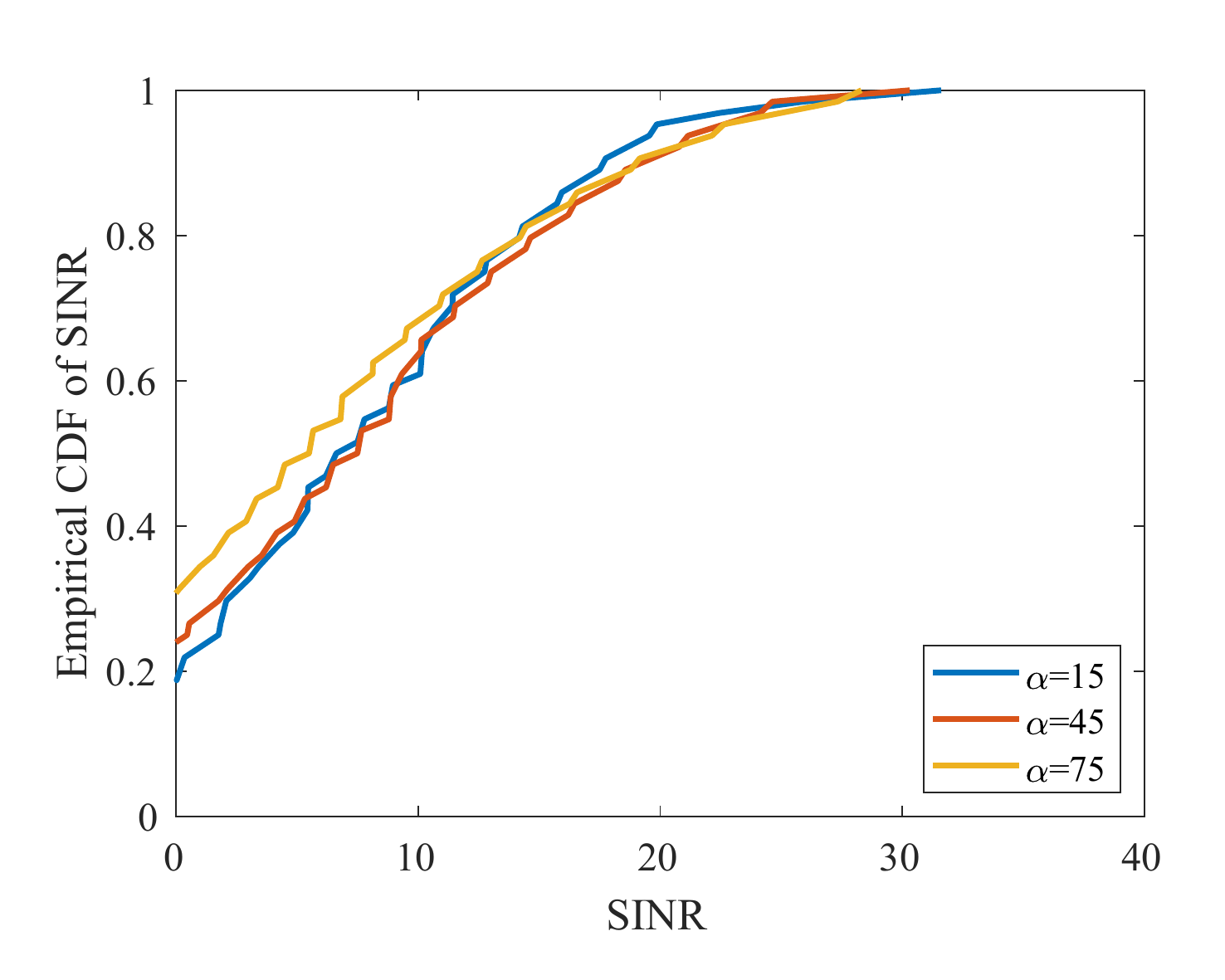}
\caption{Receivers' SINR CDF with different beamwidths} \label{sinrdisVsbeamwidth}
\end{minipage}	 \\
\vspace*{-0.25in}
\end{figure*}
\subsection{Impact of Blockage}\label{subImpactOfBlockage}

Before analyzing broadcast coverage, the impact of blockage needs to be discussed. From Fig. \ref{blockage}, we can see that one blocker will reduce the received power by 8dB, while two blockers will bring more than 25dB extra path loss in V2V communication. Due to this, we assume that the received signal power of a link that has two or more blockers is so small that it can be ignored.
\subsection{Broadcast Coverage}

As mentioned in section \ref{pes}, the coverage of a transmitter can be reflected by the number of its receivers whose SINRs are larger than a threshold $\gamma_{th}$. The results of the numbers of receivers with SINR $>$ 23dB for different beamwidths and vehicle densities are presented in Fig. \ref{density}.

The figure shows that both the beamwidth and the vehicle density influence the broadcast coverage. For beamwidth, there exists an optimal range in which the coverage is the largest or very close to the largest. The optimal range changes with the vehicle density and the associated probability of blockage. 
When the beamwidth is less than the optimal range, the number of receivers with SNR $>$ 23dB decreases as fewer nodes are pointed at by the main lobe; when the beamwidth is too large, the coverage also shrinks because the same total transmission power is spent in unnecessarily broader directions which in turn reduces the receiver-side signal power and generates more interference. In addition, the number of receivers with SINR$>$23dB is pretty small when the beamwidth is 360 degrees,  which reflects the poor result of using the omnidirectional beam pattern.

For the vehicle density, it can influence both the number of receivers and the optimal range of the beamwidth. Notice that due to the carrier sensing mechanism, the density of interferers doesn't vary with the vehicle density. The change of vehicle density mainly affects the number of potential receivers of the transmitter and the blockage probability. The median of the optimal beamwidth range increases with the vehicle density, which means for denser vehicle networks, the best strategy is to use a wider beam to cover more nearby nodes rather than using a narrower beamwidth to cover farther away nodes. Focusing on the maximum number of receivers with SINRs no less than the threshold for different vehicle densities, an interesting finding is that a sparser vehicular network tends to have a higher upper bound on the broadcast coverage. 
This is because, in a sparser network, there is a higher probability that the signal can be sent to the receivers on the adjacent lanes of the transmitter without any blockage.

To further understand what happens in the optimal beamwidth ranges, we show in Fig.~\ref{sinrdisVsbeamwidth} the empirical CDF of the top 100 receiver-side SINRs for different beamwidths when the traffic density is ``intermediate" (see Table~\ref{tab2}). 
We see that the distributions of SINR under different beamwidths are almost the same, so all the beamwidths in the optimal beamwidth range can be considered to have similar performance. 
    Fig.~\ref{density} shows that the range of optimal beamwidths tends to increase with decreasing vehicle density. Hence, it's more critical to choose the right beamwidth when the vehicle network is sparser, and, in a denser vehicle network, it tends to be easier to choose a beamwidth that falls into the optimal beamwidth range.
\begin{figure*}[!tb] 
\vspace*{-0.1in}
\centering
\figSpacing
\begin{minipage}[t]{0.3\linewidth}
\centering
\setlength{\abovecaptionskip}{0.cm}
\includegraphics[width=\textwidth]{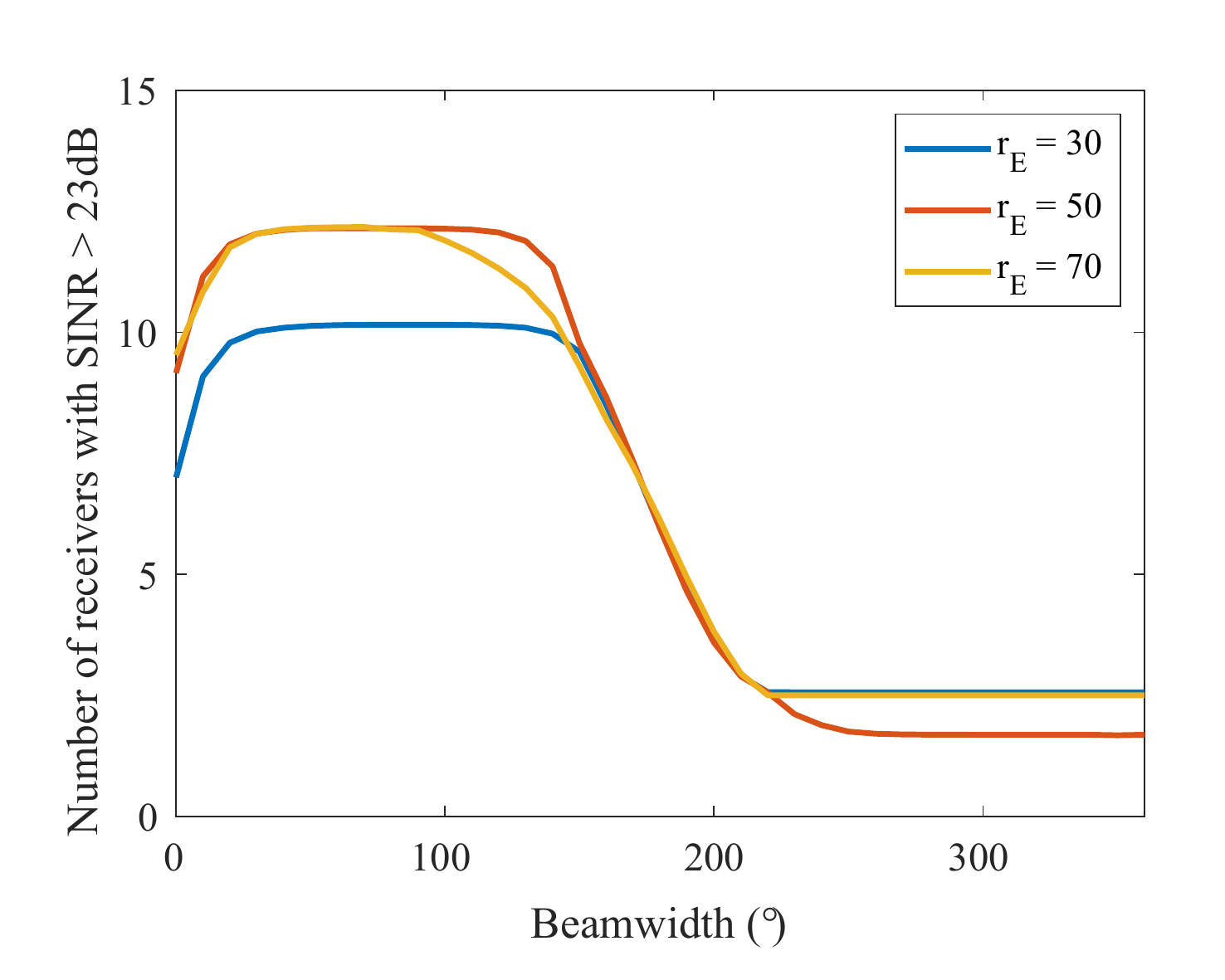}
\caption{The impact of carrier sensing range} \label{r_E}
\end{minipage}	
\figSpacing
\begin{minipage}[t]{0.3\linewidth}
\centering
\setlength{\abovecaptionskip}{0.cm}
\includegraphics[width=\textwidth]{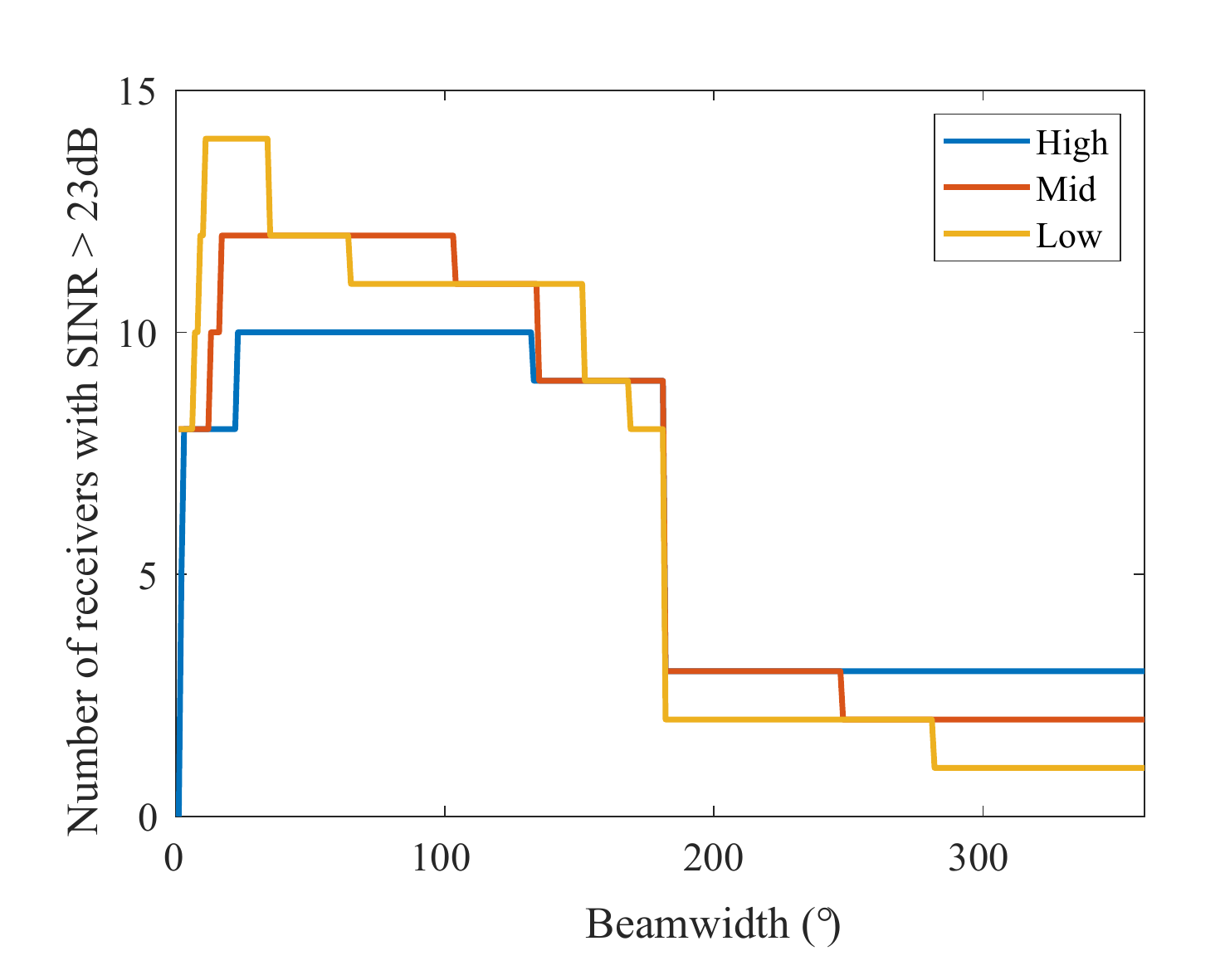}
\caption{The impacts of beamwidth and vehicle density (simulation)} \label{density_sim}
\end{minipage}	
\figSpacing
\begin{minipage}[t]{0.3\linewidth}
\centering
\setlength{\abovecaptionskip}{0.cm}
\includegraphics[width=\textwidth]{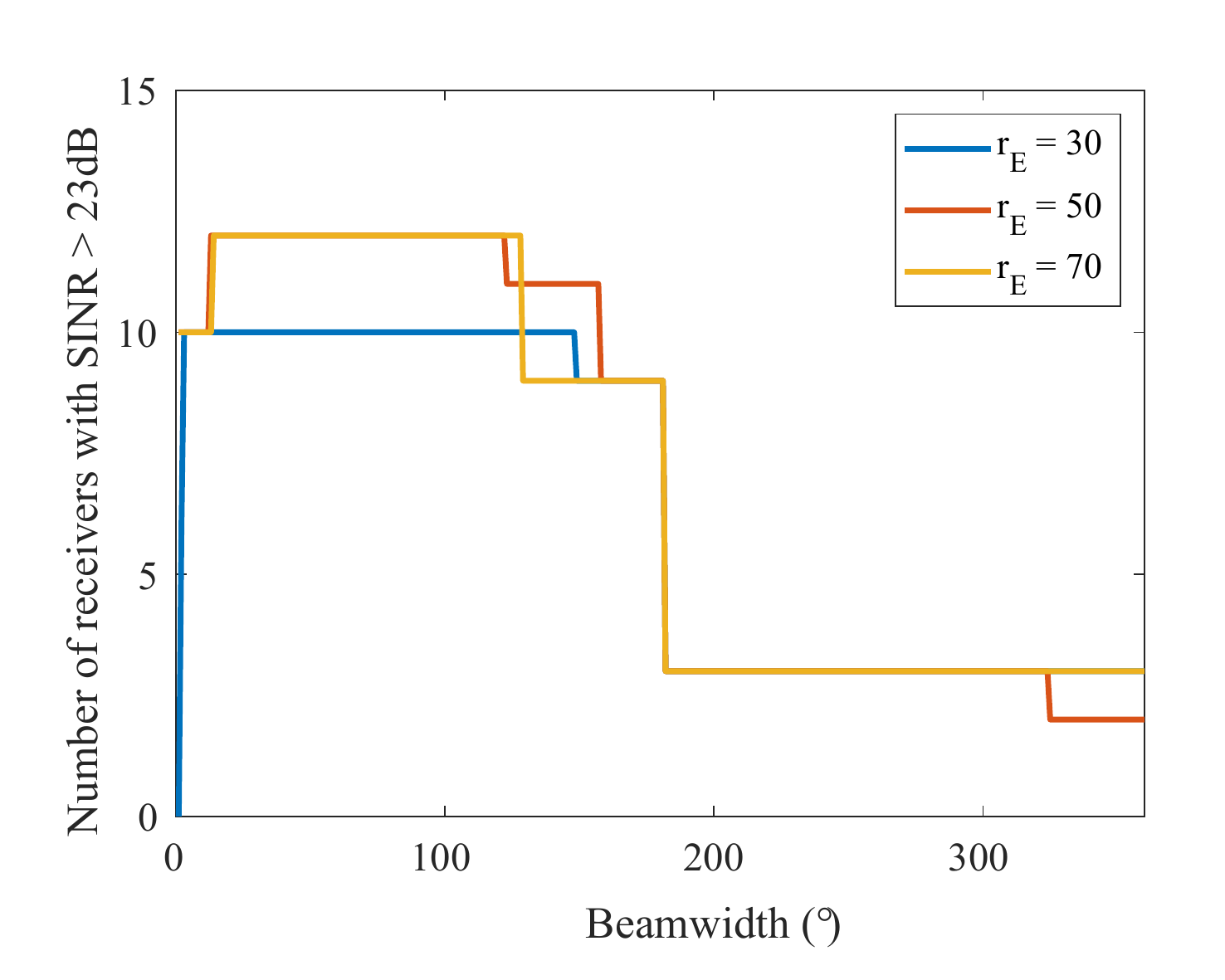}
\caption{The impact of carrier sensing range (simulation)} \label{r_E_sim}
\end{minipage}	 \\
\vspace*{-0.3in}
\end{figure*}
\subsection{Impact of Carrier Sensing Range}

The impact of carrier sensing range $r_E$ is shown in Fig. \ref{r_E}. The number of receivers with SINR no less than the given threshold varies with the carrier sensing range, and there exists an optimal value for this range. As shown by the blue line in Fig. \ref{r_E}, 
when the picked carrier sensing range is less than the optimal value, the interferers will be too close to the receiver so the SINR will decrease. However, if the picked carrier sensing range is too large, as shown by the yellow line in Fig. \ref{r_E}, the per-broadcast coverage 
doesn't increase but the density of transmitters in the networks decreases, so the overall data rate decreases. An interesting topic for future study will be to find the solutions that can decide on-the-fly the optimal carrier sensing range in real-world settings.

\subsection{Simulation Validation of Analytical Results}

To validate the analytical results, we simulate the network as specified by Tables~\ref{tab1}, \ref{tab2}, and \ref{tab3}. Fig.~\ref{density_sim} shows the impacts of beamwidth and vehicle density on broadcast coverage, and Fig.~\ref{r_E_sim} shows the impact of carrier sensing range. We see that the  analytical insight as shown by Fig.~\ref{density} is validated by Fig.~\ref{density_sim}, so is the analytical insight of Fig.~\ref{r_E} validated by Fig.~\ref{r_E_sim}. 

%% file: conclusion.tex
\section{Conclusion} \label{sec:Conclusion}
Based on high-fidelity models and industry standards, we have investigated mmWave V2V broadcast interference and coverage, 
and have studied the impacts of 
system parameters such as vehicle density and carrier sensing range. Key findings are as follows:
1) There exists an optimal beamwidth range in which the beamwidths have the similar and best performance. The range is influenced by the vehicle density, and it is smaller when the network is sparser; this implies that optimal beamwidth selection is more important in a sparser network, and it is relatively easier to find the beamwidth that maximizes the broadcast coverage in a denser network. 
2) Vehicle density influences the median of the optimal beamwidths, in addition to their range. For a denser network, it's better to use a larger beamwidth to cover more nearby nodes while, for a sparser network, choosing a narrower beamwidth to cover more far-away nodes is preferred. The density also impacts the upper bound on broadcast coverage. With less blockage, sparser vehicle networks tend to have higher coverage upper bound. 
3) Interference control is important for mmWave V2V broadcast, and the optimal selection of carrier sensing range is important as it impacts inter-vehicle interference as well as the concurrency and throughput of V2V broadcast.
    These findings provide unique insight into optimal beamwidth selection in mmWave V2V networks, and they also demonstrate the importance of interference control in mmWave V2V broadcast. 
An interesting future work is to develop mmWave V2V broadcast protocols based on these findings.
\vspace*{-0.04in}